\documentstyle[12pt]{article}

\def\hybrid{\topmargin -20pt	\oddsidemargin 0pt
	\headheight 0pt	\headsep 0pt
        \textwidth 6.35in
        \textheight 9.65in
	\marginparwidth .875in
	\parskip 5pt plus 1pt	\jot = 1.5ex}

 \catcode`\@=11

\@addtoreset{equation}{subsection}
\@addtoreset{footnote}{section}
\def\theequation{\thesubsection.\arabic{equation}}

\newtoks\@stequation

\def\subequations{\refstepcounter{equation}%
  \edef\@savedequation{\the\c@equation}%
  \@stequation=\expandafter{\theequation}
  \edef\@savedtheequation{\the\@stequation}
  \edef\oldtheequation{\theequation}%
  \setcounter{equation}{0}%
  \def\theequation{\oldtheequation\alph{equation}}}

\def\endsubequations{\setcounter{equation}{\@savedequation}%
  \@stequation=\expandafter{\@savedtheequation}%
  \edef\theequation{\the\@stequation}\global\@ignoretrue
  \vspace*{-12pt} \\}

\def\pr{Phys. Rev. \/}
\def\np{Nucl. Phys. \/}
\def\pl{Phys. Lett. \/}

\hybrid
\def\baselinestretch{1.2}
\catcode`\@=11
\newcount\hour
\newcount\minute
\newtoks\amorpm
\hour=\time\divide\hour by60
\minute=\time{\multiply\hour by60 \global\advance\minute by-\hour}
\edef\standardtime{{\ifnum\hour<12 \global\amorpm={am}%
	\else\global\amorpm={pm}\advance\hour by-12 \fi
	\ifnum\hour=0 \hour=12 \fi
	\number\hour:\ifnum\minute<10 0\fi\number\minute\the\amorpm}}
\edef\militarytime{\number\hour:\ifnum\minute<10 0\fi\number\minute}

\def\draftlabel#1{{\@bsphack\if@filesw {\let\thepage\relax
   \xdef\@gtempa{\write\@auxout{\string
      \newlabel{#1}{{\@currentlabel}{\thepage}}}}}\@gtempa
   \if@nobreak \ifvmode\nobreak\fi\fi\fi\@esphack}
	\gdef\@eqnlabel{#1}}
\def\@eqnlabel{}
\def\@vacuum{}
\def\draftmarginnote#1{\marginpar{\raggedright\scriptsize\tt#1}}

\def\draft{\oddsidemargin -.5truein
	\def\@oddfoot{\sl preliminary draft \hfil
	\rm\thepage\hfil\sl\today\quad\militarytime}
	\let\@evenfoot\@oddfoot	\overfullrule 3pt
	\let\label=\draftlabel
	\let\marginnote=\draftmarginnote
   \def\@eqnnum{(\theequation)\rlap{\kern\marginparsep\tt\@eqnlabel}%
\global\let\@eqnlabel\@vacuum}  }

\catcode`\@=11
\def\section{\@startsection {section}{1}{0pt}{-3.5ex plus -1ex minus
 -.2ex}{2.3ex plus .2ex}{\raggedright\large\bf}}
\catcode`\@=12
\newskip\humongous \humongous=0pt plus 1000pt minus 1000pt

\newif\ifdtup

\def\be{\begin{equation}}
\def\ee{\end{equation}}
\def\ba{\begin{eqnarray}}
\def\ea{\end{eqnarray}}
\def\bs{\begin{subequations}}
\def\es{\end{subequations}}

\def\th{\theta}

\def\tb{\bar{\theta}}

\def\t{\tau}
\def\tb{\bar\tau}
\def\im{{\rm Im}\tau}

\def\t{\tau}

\def\R{{\cal{R}}}
\def\Q{{\cal{Q}}}

\def\dslash{{\partial\hspace{-.22cm}/}}
\def\hslash{{\rm H}\hspace{-.28cm}/}
\begin{document}
\renewcommand{\theequation}{\thesection.\arabic{equation}}
\newcommand{\beq}{\begin{equation}}
\newcommand{\eeq}[1]{\label{#1}\end{equation}}
\newcommand{\ber}{\begin{eqnarray}}
\newcommand{\eer}[1]{\label{#1}\end{eqnarray}}
\begin{titlepage}
\begin{center}

\hfill CERN-TH/95-222\\
\hfill LPTENS-95/40\\
\hfill hep-th/9509017\\

\vskip .2in

{\large \bf
One-Loop Calculation of Coupling Constants in IR regulated String
Theory\footnote{To appear in the proceedings of the Trieste Spring
School and Workshop on String Theory, 27 March-7 April 1995.}}
\vskip .4in

{\bf Elias Kiritsis and Costas Kounnas\footnote{On leave from Ecole
Normale Sup\'erieure, 24 rue Lhomond, F-75231, Paris, Cedex 05,
FRANCE.}}\\
\vskip
 .3in

{\em Theory Division, CERN,\\ CH-1211,
Geneva 23, SWITZERLAND} \\

\vskip .3in

\end{center}

\vskip .2in

\begin{center} {\bf ABSTRACT } \end{center}
\begin{quotation}\noindent

Exact Superstring solutions are constructed moving in 4-D space-time
with positive curvature
and non-trivial dilaton and antisymmetric tensor fields. The full
spectrum
of string excitations is derived as a function of moduli fields
$T^{i}$ and the scale $\mu^2=1/(k+2)$ which induced by the non-zero
background fields.
The spectrum of string excitations has a non-zero mass gap $\mu^2$
and in the weak curvature limit ($\mu$ small) $\mu^2$ plays the role
of a well defined infrared regulator, consistent with modular
invariance, gauge invariance,
supersymmetry and chirality.\\
 The effects of a covariantly constant chomo-magnetic field $B$ as
well as additional curvature can be derived exactly up to one
string-loop level. Thus,  the one-loop corrections to
all couplings (gravitational, gauge and Yukawas) are
unambiguously computed and are finite both in the Ultra-Violet and
the Infra-Red regime.
These corrections are necessary for quantitative string
superunification predictions at low energies. The one-loop
corrections to the couplings are also found to satisfy
Infrared Flow Equations.

\end{quotation}
\vskip 1.0cm
CERN-TH/95-222 \\
August 1995\\
\end{titlepage}
\vfill
\eject
\def\baselinestretch{1.2}
\baselineskip 16 pt
\noindent
\section{Introduction}
\setcounter{equation}{0}

Four-dimensional superstring solutions in a flat background
\cite{cand}-\cite{gepner}
 define, at  low energy, effective supergravity theories
\cite{effcl}-\cite{ant}.
 A class of them successfully extends
the validity of the standard model up to the string scale,
$M_{str}$.
 The first main property of superstrings is that they are
ultraviolet-finite theories (at least perturbatively). Their
second important
property is that  they unify gravity with all other interactions.
This unification does not include  only the gauge interactions, but
also the  Yukawa  ones as well as the interactions among the scalars.
This String Unification happens  at  large energy scales
$E_t\sim {\cal O}(M_{str})\sim 10^{17}~$GeV. At this energy scale,
however,
the first excited string states become important and thus the whole
effective low energy field theory picture breaks
down \cite{n4kounnas}-\cite{kktopol}. Indeed, the
effective field theory of strings is valid only for  $E_t \ll
M_{str}$ by means of
the ${\cal O}(E_t/M_{str})^2$ expansion. It is then necessary to
evolve the String Unification predictions to a lower scale $M_U <
M_{str}$ where
the  effective field theory picture makes sense. Then, at $M_U$, any
string solution provides  non-trivial relations between  the gauge
and Yukawa couplings, which can be written as

\begin{equation}
\frac{k_i}{\alpha_i(M_U)}={1\over \alpha_{str}}+\Delta_{i}(M_U).
\label{shu}
\end{equation}

The above relation looks very similar to the well-known unification
condition
in Supersymmetric Grand Unified Theories (SuSy-GUTs) where the
unification scale is about $M_U\sim 10^{16}~$GeV and
$\Delta_{i}(M_U)=0$ in the ${\bar {DR}}$ renormalization scheme; in
SuSy-GUTs the normalization constants $k_i$ are fixed $only$ for the
gauge couplings ($k_1=k_2=k_3=1$, $k_{em}=\frac{3}{8}$), but there
are
no relations among gauge and Yukawa couplings at all. In string
effective theories, however, the normalization constants ($k_i$) are
known for both gauge and Yukawa interactions. Furthermore,
$\Delta_{i}(M_U)$ are calculable $finite$ quantities for any
particular string solution. Thus, the predictability of a given
string solution is extended for all low energy coupling constants
${\alpha_i(M_Z)}$ once the string-induced corrections
$\Delta_{i}(M_U)$ are determined.

It turns out that
$\Delta_{i}(M_U)$ are non-trivial functions of the vacuum
expectation values  of  some gauge singlet fields
\cite{moduli,dfkz,ant,n2},
$~~\langle T_A\rangle =t_A$, the so-called moduli (the moduli fields
are flat directions at the string classical level and they remain
flat in
string perturbation theory, in the exact supersymmetry limit).
The $\Delta_{i}(t_A)$ are target space duality invariant functions,
which depend on the particular
string ground state. $\beta$-functions in string theory were
calculated first in \cite{min}. Several results for
$\Delta_{i}$ exist by now \cite{moduli,dfkz,ant,n2} in the exact
supersymmetric limit in many string
solutions based on orbifold
\cite{orbifold} and fermionic constructions \cite{abk4d}.
As we will see later $\Delta_{i}$
are, in principle,   well defined
calculable quantities once we perform our calculations  at the string
level where all interactions including gravity are  consistently
defined.
Although at this stage we do not know the details of supersymmetry
breaking
we should stress here that for dimensionless couplings in four
dimensions
choosing the IR (low energy) scale above the threshold of
supersymmetric partners
the one-loop corrections to these couplings is exact because soft
breaking terms do not affect such renormalizations.
For soft breaking parameters only qualitative results exist up to now
although this is a subject of intensive study.

The main obstruction in determining the exact form of the string
radiative corrections $\Delta_{i}(M_U)$ is related to the
infrared divergences of on-shell calculations in string theory.
In field
theory, we
can avoid this problem using off-shell calculations. In first
quantized string theory we cannot do that since we do not know how to
go off-shell.
Even in field theory there are problems in defining an infrared
regulator for chiral fermions especially in the presence of
spacetime supersymmetry.

The idea is to modify slightly the ground state of interest in string
theory
so that it develops a mass gap. It is known already in field theory
that a space of negative curvature provides bosonic fields (scalars,
vectors etc.) with such a mass gap.
The case of fermions (especially chiral ones) though is more subtle.
We will see however that string theory contains the fields (namely
the antisymmetric tensor) which when they acquire some suitable
expectation values they can provide a mass gap for (chiral) fermions.

Let us indicate here how an expectation value for the dilaton can
give masses to bosonic fields.
The dilaton couples generically to (massless) bosonic fields in a
universal fashion:

\be
S_{T}=\int e^{-2\Phi}\partial_{\mu}T\partial^{\mu}T
\ee
where we considered the case of a scalar field $T$.
To find the spectrum of the fluctuations of T we have to define
$\tilde T=e^{-\Phi}T$ so that the action becomes
\be
S_{T}=\int \partial_{\mu}\tilde T\partial^{\mu}\tilde
T+\left[\partial_{\mu}\Phi\partial^{\mu}\Phi-\partial_{\mu}\partial^{\mu
}\Phi\right]\tilde T^2
\ee
It is obvious that if $\langle \Phi\rangle =Q_{\mu}x^{\mu}$ then the
scalar
$\tilde T$ acquires a mass$^2 ~Q_{\mu}Q^{\mu}$ which is positive when
$Q_{\mu}$ is spacelike\footnote{This was observed in \cite{aben} with
$Q_{\mu}$ being timelike.} .
Similar remarks apply to higher spin bosonic fields.
This mechanism via the dilaton cannot give masses to fermions since
the extra shift obtained by the redefinition is a total divergence.

Consider a chiral fermion with its universal coupling to the
antisymmetric tensor:

\be
S_{\psi}=\int \bar \psi [{\buildrel{\leftrightarrow}\over
\dslash}+{\hslash}]\psi
\ee
where $H_{\mu}={\epsilon_{\mu}}^{\nu\rho\sigma}H_{\nu\rho\sigma}$ is
th dual
of the field stregth of the antisymmetric tensor.
If $\langle H_{\mu}\rangle =Q_{\mu}$ then the Dirac operator acquires
a mass gap proportional to $Q_{\mu}Q^{\mu}$.

Thus we need to find exact string ground states (CFTs) which
implement the mechanism sketched above.

In particular we would like our background to have the following
properties:

{\bf 1.} The string spectrum must have a mass gap $\mu^2$.
          In particular, chiral fermions should be regulated
consistently.

{\bf 2.} We should be able to take the limit $\mu^2\to 0$.

{\bf 3.} It should have $c=(6,4)$ so that it can be coupled to any
         internal CFT with $c=(9,22)$.

{\bf 4.} It should preserve as many spacetime supersymmetries of the
original theory, as possible.

{\bf 5.} We should be able to calculate the regulated quantities
relevant for
the effective field theory.

{\bf 6.} Vertices for spacetime fields (like $F_{\mu\nu}^{a}$) should
be
          well defined operators on the world-sheet.

{\bf 7.} The theory should be modular invariant (which guarantees the
absence
of anomalies).

{\bf 8.} Such a regularization should be possible also at the
effective field theory level. In this way, calculations in the
fundamental theory can be matched
without any ambiguity to those of the effective field theory.

CFTs with the properties above employ special four-dimensional spaces
as
superconformal building blocks with
${\hat c}=4$ and $N=4$ superconformal
symmetry \cite{n4kounnas,worm}. The full spectrum of string
excitations for the superstring solutions based on those
four-dimensional subspaces, can be derived using the techniques
developed in \cite{worm}. The spectrum does have a mass gap, which is
proportional
to the curvature of the non-trivial  four-dimensional spacetime.
Comparing the spectrum in  a flat background with that in curved
space we observe a shifting of all massless states by an amount
proportional to the spacetime curvature, $\Delta m^2=Q^2=\mu^2$,
where $Q$ is the Liouville background charge and $\mu$ is the IR
cutoff. What is
also interesting is that the shifted spectrum in curved space is
equal for bosons and fermions due to the existence of a new
space-time supersymmetry defined in curved spacetime
\cite{n4kounnas,worm}. Therefore, our curved spacetime
infrared
regularization is consistent with supersymmetry and can be
used either in field theory or string theory.

Most of the work presented here has already appeared in
\cite{trieste,magn}
We present also some new results in section 4.

\section{The IR regulated String Theory}
\setcounter{equation}{0}

We will choose the 4-D CFT which will replace flat space to be
the $SU(2)_{k}\otimes R_{Q}$ model.
It contains a non-compact direction with a linear dilaton
$\Phi=Qx^{0}$
as well as the $SU(2)_{k}$ WZW model.
Q is related to $k$ as $Q=1/\sqrt{k+2}$ so that the CFT has the same
central charge as flat space.
We will define $\mu^2=1/(k+2)$, $\mu$ is directly related to the mass
gap of the regulated theory.
The GSO projection couples the SU(2) spin with the spacetime helicity
\cite{trieste}.
This effectively projects out the half-integral spins and replaces
$SU(2)$ with $SO(3)$. $k$ should be an even positive integer for
consistency.
For any ground state of the heterotic string with $N<4$ spacetime
supersymmetry
the regulated vaccum amplitude turns out to be

\be
Z(\mu)={1\over V(\mu)}\Gamma_{0}(\mu)Z_{0}\label{21}
\ee
where $V(\mu)=1/8\pi\mu^3$ is the volume of the nontrivial background
and $Z_{0}$ is the vacuum amplitude for the unregulated theory, which
can be written as
\be
Z_{0}(\t,\tb)={1\over \im|\eta|^4}\sum_{a,b=0}^{1}{\th[^a_b]\over
\eta}
C[^a_b](\t,\tb)
\label{22}
\ee
where we have separated the generic 4-d contribution. The factor
$C[^a_b]$ is the Trace in the $(^a_b)$ sector of the internal CFT.
Finally, $\Gamma_{0}(\mu)$ is proportional to the $SO(3)_{k}$
partition function
\be
\Gamma_{0}(\mu)={1\over 2}[({\rm Im}\tau)^{\frac{1}{2}}
\eta(\tau){\bar
\eta}({\bar \tau})]^{3}~~\sum_{a,b=0}^{1}e^{-i\pi
kab/2}\sum_{l=0}^{k}e^{i\pi
bl}\chi_{l}(\t)\bar \chi_{(1-2a)l+ak}(\tb)
\label{23}
\ee
where $\chi_{l}$ are the standard $SU(2)_{k}$ characters.
We have also the correct limit $Z(\mu)\to Z_{0}$ as $\mu\to 0$.

There is a simple expression for $\Gamma_{0}(\mu)$
\be
\Gamma_{0}(\mu)=-{1\over 2\pi}X'(\mu)
\label{24}
\ee
where prime stands for derivative with respect to $\mu^2$
and
\be
X(\mu)={1\over 2\mu}\sum_{m,n\in
Z}e^{i\pi(m+n+mn)}\;\exp\left[-{\pi\over
4\mu^2\im}|m-n\t|^2\right]=\sqrt{\im}\sum_{m,n\in Z}e^{i\pi
n}q^{{1\over
4}Q^{2}_{L}}\bar q^{{1\over 4} Q^{2}_{R}}
\label{25}
\ee
with
\be
Q_{L}=2\mu\left(m-{n+1\over 2}\right)+{n\over
2\mu}\;\;,\;\;Q_{R}=2\mu\left(m-{n+1\over 2}\right)-{n\over 2\mu}
\label{26}
\ee
It can be also written in terms of the usual torroidal sum:
\be
X(\mu)=Z_{T}(\mu)-Z_{T}(2\mu)
\ee
\be
Z_{T}(\mu)=Z_{T}(1/\mu)=\sqrt{\im}\sum_{m,n\in Z}q^{{1\over
4}\left(m\mu+n/\mu\right)^2}
\bar q^{{1\over 4}\left(m\mu-n/\mu\right)^2}\label{37}
\ee
Note that $X(\mu)$ is modular invariant.

The leading infrared behaviour can be read from (\ref{24}),
(\ref{25})
to be
\be
Z(\mu)\to \sqrt{\im}e^{-\pi\im\mu^2}
\label{27}
\ee
as $\im\to \infty$ that indicates explicitly the presence of the mass
gap.

More details on this theory can be found in \cite{trieste,magn}.

\section{Non-zero ${\bf F^{a}_{\mu\nu}}$ and  $
R_{\mu\nu}^{\rho\sigma}$ Background in Superstrings}
\setcounter{equation}{0}

As mentioned in the introduction in order to calculate the
renormalization of the effective couplings we need to turn on
backgrounds for gauge and gravitational fields.
Thus our aim is to define the  deformation of the two-dimensional
superconformal theory  which corresponds to a non-zero field strength
$F^{a}_{\mu\nu}$ and $R_{\mu\nu\rho\sigma}$
background\footnote{Magnetic backgrounds in closed string theory have
been also discussed in \cite{bk,rt}.}
and find the integrated  one-loop
partition function $ Z(\mu,F,\R)$,  where $F$ is by the
magnitude
of the field strength,
$F^2 \equiv \langle F^{a}_{\mu\nu}F_{a}^{\mu\nu}\rangle$ and $\R$ is
that of the curvature,  $\langle
R_{\mu\nu\rho\sigma}R^{\mu\nu\rho\sigma}\rangle=\R^2$.

\begin{equation}
Z[\mu,F_i,\R]=\frac{1}{V(\mu)} \int_{\cal F}
\frac{ d\tau d{\bar\tau} }{ ({\rm Im}\tau)^2 }
Z[\mu,F_i,\R;\tau,{\bar\tau}]
\label{intpart}
\end{equation}
The index $i$ labels different simple or $U(1)$ factors of the gauge
group of the ground state.

 In flat space, a small non-zero  $F_{\mu\nu}^a$ background gives
rise
to an infinitesimal deformation  of the 2-d $\sigma$-model action
given by,
\begin{equation}
\Delta S^{2d}(F^{(4)})=\int dzd{\bar z}\;F_{\mu\nu}^a[x^{\mu}
\partial_z x^{\nu}+\psi^{\mu}\psi^{\nu}]{\bar J}_a
\label{fdef}
\end{equation}
Observe that for $F^a_{\mu\nu}$ constant (constant magnetic field),
the left moving operator $[x^{\mu} \partial_z
x^{\nu}+\psi^{\mu}\psi^{\nu}]$ is not a well-defined $(1,0)$ operator
on the world sheet. Even though  the right moving Kac-Moody current
${\bar J}_a$ is a well-defined $(0,1)$ operator, the total
deformation
is not integrable in flat space. Indeed, the 2-d $\sigma$-model
$\beta$-functions are not satisfied in the presence of a constant
magnetic field. This follows from the fact that there is a
$non$-$trivial$ $back$-$reaction$ on the gravitational background due
the non-zero
magnetic field.

The important property of our non-trivial spacetime background is
that we can solve exactly for the
back-reaction. First observe that the deformation that
corresponds to a constant magnetic field
$B_i^a=\epsilon_{oijk}F_a^{ik}$ is a well-defined
(1,1) integrable deformation, which breaks the $(2,2)$ superconformal
invariance but preserves the $(1,0)$ world-sheet supersymmetry:
\begin{equation}
\Delta S^{2d}(W^{(4)}_k)=\int dzd{\bar
z}\;B^a_i[I^i+\frac{1}{2}\epsilon^{ijk}\psi_{j}\psi_{k}]{\bar J}_a
\label{fdef1}
\end{equation}
where $I^i$ is anyone of the $SU(2)_{k}$ currents.
The deformed partition function is not zero due to the breaking of
$(2,2)$ supersymmetry.
It can be shown that this is  the correct replacement of the Lorentz
current in the flat case \cite{trieste}.
More details on the physics of such magnetic fields can be found in
\cite{magn}.

The $\cal{R}$ perturbation is
\be
\Delta S({\cal{R}})=\int dzd\bar
z\;{\cal{R}}\left[I^{3}+\psi^{1}\psi^{2}\right]
\bar I^{3}
\ee
This turns on a Riemann tensor with constant scalar curvature equal
to
$6{\cal{R}}$ as well as an antisymmetric tensor and dilaton.

Due to the rotation invariance in $S^3$ we can choose
$B_i^a=F\delta_i^3$ without loss of generality. The vector
$B^{a}_{i}$
indicates the
direction in the gauge group space of the right-moving affine
currents.

The moduli space of the $F$ deformation is then given by the
$SO(1,n)/SO(n)$ Lorentzian-lattice boosts with $n$ being  the rank
of the right-moving gauge group. We therefore conclude that the
desired partition function $Z(\mu,F_i,\R=0)$  is given in terms
of the moduli of the $\Gamma(1,n)$ lorentzian lattice. The constant
gravitational background
$R^{ij}_{kl}=\R\epsilon^{3ij}\epsilon_{3kl}$ can also be
included exactly
by an extra boost, in which case the lattice becomes
$\Gamma(1,n+1)$.

Let us denote by  $\Q$ the fermionic lattice momenta associated to
the left-moving $U(1)$ current $\partial H=\psi^1\psi^2$, by
$J^{3},\bar J^{3}$
the zero mode of the respective (left or right) $SU(2)_{k}$ current,
by $\bar \Q_i$ the zero mode of a Cartan generator of the left moving
current algebra generating a simple or U(1) component of the gauge
group.
The currents are normalized so that
$J^{a}_{i}(z)J^{b}_{j}(0)=k_{i}\delta_{ij}\delta^{ab}/2z^2+...$.
Note that this fixes the normalization of the Casimirs.

In terms of these charges the undeformed partition
function can be written as
\be
Tr[\exp[-2\pi \rm{Im}\tau (L_{0}+\bar L_{0})
+2\pi i\rm{Re}\tau (L_{0}-\bar L_{0})]]
\end{equation}
where
\be
{\rm L}_{0}={1\over 2}\Q^2+{(J^3)^2\over k}+\cdots\;,\;\bar {\rm
L}_{0}=\sum_{i}{\bar
\Q_i^2\over k_{i}}+{(\bar J^3)^2\over k}+\cdots
\ee
where the dots stand for operators that do not involve $J^3,\bar
J^3,\Q,\bar \Q_i$.

The (1,1) perturbation that turns on  constant gauge field strengths
$F_i$, $i=1,2,...,n$ as well
as a constant curvature $\R$ background
produces an O(1,1+n) 2-parameter boost in $O(2,2)$, acting on the
charge lattice
above.

\be
\delta L_{0}=\delta \bar L_{0}=(\Q+J^3)(\R \bar J^3+F_{i}\bar Q_{i})+
\ee
$$+{-1+\sqrt{1+(k+2)(k_{i}F_{i}^2+k\R^2)}\over 2}\left[
{(\Q+J^3)^2\over k+2}+{(F_{i}\bar Q_{i}+\R\bar J^3)^2\over
k_{i}F_{i}^2+k\R^2}
\right]
$$

The first term is the standard perturbation while the second term is
the back-reaction necessary for conformal and modular invariance.
Expanding the partition function in a power series in $F_i,\R$
\begin{equation}
Z(\mu,F,\R)=\sum_{n_i,m=0}^{\infty}\prod_{i=1}^{n}F_i^{n_i}
\R^{m}Z_{n_i,m}(\mu)
\end{equation}
we can extract the integrated correlators $Z_{n_i,m}=\langle
\prod_{i=1}^{n}F_i^{n_i} R^m\rangle$.
\bs
\label{formu}
\be
\langle F_{i}\rangle =-4\pi {\rm Im}\tau \langle
(\Q+J^3)\rangle\langle
\bar Q_{i}\rangle
\ee
\be
\langle \R\rangle =-4\pi {\rm Im}\tau \langle (\Q+J^3)\rangle\langle
\bar J^3\rangle
\ee
\be
\langle F_{i}^2\rangle =8\pi^2{\rm Im}\tau^2\left[ \langle
(\Q+J^3)^2\rangle
-{(k+2)\over 8\pi\im}\right]\left[\langle (\bar
Q_{i})^2\rangle-{k_{i}
\over 8\pi{\rm Im}\tau}\right]-{k_{i}(k+2)\over 8}
\label{F2}
\ee
\be
\langle \R^2\rangle =8\pi^2{\rm Im}\tau^2\left[ \langle
(\Q+J^3)^2\rangle
-{k+2\over 8\pi\im}\right]\left[\langle (\bar J^3)^2\rangle-{k
\over 8\pi{\rm Im}\tau}\right]-{k(k+2)\over 8}
\ee
\be
\langle \R F_{i}\rangle = 16\pi^2{\rm Im}\tau^2 \langle \bar J^3\bar
Q_{i}\rangle
\left[\langle (\Q+J^3)^2\rangle-{k+2\over 8\pi{\rm Im}\tau}\right]
\ee
\be
\langle F_{i}F_{j}\rangle = 16\pi^2{\rm Im}\tau^2 \langle \bar
Q_{i}\bar Q_{j}\rangle
\left[\langle (\Q+J^3)^2\rangle-{k+2\over 8\pi{\rm Im}\tau}\right]
\ee
\es
where we should always remember that $k+2=1/\mu^2$.

For Supersymmetric ground states we have simplifications
\be
\langle F_{i}^2\rangle_{SUSY}=8\pi^2{\rm
Im\tau}^2\langle\Q^2\rangle\left[
\langle (\bar Q_{i})^2\rangle-{k_{i}\over 8\pi{\rm Im}\tau}\right]
\ee
\be
\langle \R^2\rangle_{SUSY}=8\pi^2{\rm
Im\tau}^2\langle\Q^2\rangle\left[
\langle (\bar J^3)^2\rangle-{k\over 8\pi{\rm Im}\tau}\right]
\ee

Renormalizations of higher terms can be easily computed.
We give here the expression for an $F_{i}^4$ term,

$$
\langle F_{i}^4\rangle={(4\pi{\rm Im}\tau)^4\over 24}\langle\left[
(\Q+J^3)^4  (\bar Q_{i})^4-{3\over 4\pi{\rm Im}\tau}(\Q+J^3)^2(\bar
Q_{i})^2
\left((k_{i}(\Q+J^3)^2+\right.\right.
$$
\be
+\left.(k+2)(\bar Q_{i})^2\right)
+{3\over 4(4\pi{\rm Im}\tau)^2}\left[k_{i}(\Q+J^3)^2+(k+2)(\bar
Q_{i})^2\right]^2-\label{F4}
\ee
$$
-\left.{3k_{i}(k+2)\over 2(4\pi{\rm
Im}\tau)^3}\left[[k_{i}(\Q+J^3)^2+(k+2)(\bar
Q_{i})^2\right]
\right]\rangle
$$

The charge $\Q$ in the above formulae acts on the helicity
$\vartheta$-function
$\vartheta\left[^{\alpha}_{\beta}\right](\tau,v)$ as
differentiation with respect to $v$ divide by $2\pi i$.
The charges $\bar Q_{i}$ act also as $v$ derivatives on the
respective characters of the current algebra.
$J^3,\bar J^3$ act in the level-$k$ $\vartheta$-function present in
$SO(3)_{k}$ partition function (due to the parafermionic
decomposition).

\section{One-loop Corrections to the Coupling Constants}
\setcounter{equation}{0}

We now focus on  the one-loop correction to the
gauge
couplings.
Bearing anomalous U(1)'s we can immediately see from (\ref{formu})
that
$\langle F_{i}\rangle =0$ and $\langle F_{i}F_{j}\rangle =0$ for
$i\not= j$.
The conventionally normalized one-loop correction is
\be
{16\pi^2\over g^2_i}|_{1-loop}=-{1\over (2\pi)^2}\int_{\cal
F}{d^2\t\over \im^2}\langle F_{i}^2\rangle
\label{121}
\ee

Putting everything together we obtain

$${16\pi^2\over g_{i}^2}|_{1-loop} =-{i\over \pi^2 V(\mu)}\int_{\cal
F}{d^2\t
\over \im |\eta|^4}
\sum_{a,b=0}^{1}\left[X'(\mu)\partial_{\t}\left({\th[^a_b]\over
\eta}\right)+{1\over 6\mu^2}
\dot X'(\mu){\th[^a_b]\over \eta}\right]\times
$$
\be
\times Tr^{I}_{a,b}\left[
\langle (\bar J^{i})^2\rangle-{k_{i}\over 8\pi{\rm Im}\tau}\right]-
-{k_{i}\over 64\pi^3\mu^2 V(\mu)}\int_{\cal F}{d^2\t\over \im^2}
X'(\mu)Z_{0}\label{35}
\ee
where dot stands for derivative with respect to $\t$ and
$Tr^{I}_{ab}$ stands for the trace in the $(^a_b)$ sector of the
internal CFT.
Eq. (\ref{35}) is valid also for non-supersymmetric ground states.

When we have $N\geq 1$ supersymmetry it simplifies to
\be
{16\pi^2\over g_{i}^2}|^{SUSY}_{1-loop} =-{i\over \pi^2
V(\mu)}\int_{\cal F}{d^2\t
\over \im |\eta|^4}
\sum_{a,b=0}^{1}\left[X'(\mu){\partial_{\t}\th[^a_b]\over
\eta}\right]Tr^{I}_{a,b}\left[
\langle (\bar J^{i})^2\rangle-{k_{i}\over 8\pi{\rm Im}\tau}\right]
\label{511}
\ee
The general formula (\ref{35}) can be split in the following way
\be
{16\pi^2\over g_{i}^2}|_{1-loop}=I_{1}+I_{2}+I_{3}\label{40}
\ee

\be
I_{1}=-{ i\over \pi^2 V(\mu)}\int_{\cal F}{d^2\t\over
\im|\eta|^4}X'(\mu)\sum_{a,b=0}^{1}
 \partial_{\t}\left({\th[^a_b]\over
\eta}\right)Tr^{I}_{a,b}\left[\langle (\bar
J^{i})^2\rangle-{k_{i}\over 8\pi{\rm Im}\tau}\right]
\label{41}
\ee

\be
I_{2}=-{i\over 6\pi^2\mu^2V(\mu)}\int_{\cal F}{d^2\t\over
\im|\eta|^4}\dot
X'(\mu)\sum_{a,b}^{1}
{\th[^a_b]\over \eta}Tr^{I}_{a,b}\left[\langle (\bar
J^{i})^2\rangle-{k_{i}\over
8\pi{\rm Im}\tau}\right]
\label{42}
\ee

\be
I_{3}=-{k_{i}\over 64\pi^3\mu^2 V(\mu)}\int_{\cal F}{d^2\t\over
\im^2}
X'(\mu)Z_{0}\label{43}
\ee
All the integrands are separately modular invariant.
The universal term in $I_{1}$ is due to an axion tadpole. $I_{3}$ is
the contribution of a dilaton tadpole. $I_{2}$ are extra helicity
contributions
due to the curved background.
Moreover $I_{2},I_{3}$ have power IR divergences which reflect
quadratic divergences in the effective field theory.
$I_{2},I_{3}$ are zero for supersymmetric ground states.

We will now analyse the contribution of the massless sector to the
one-loop corrections.
Since
\be
-{1\over i\pi }\partial_{\t}\left({\th[^a_b]\over \eta}\right)\to
(-1)^F\left({1\over 12}-\chi^2\right)
\label{444}
\ee
where $\chi$ is the helicity of a state,
we obtain
\be
I_{1}^{massless}=-{1\over \pi}Str\left[Q_{i}^2\left({1\over
12}-\chi^2\right)\right]{\bf J}_{1}(\mu)
+{k_{i}\over 8\pi^2}Str\left[{1\over 12}-\chi^2\right]{\bf
J}_{2}(\mu)
\label{445}
\ee
\be
I_{2}^{massless}=-{1\over 12\pi^2\mu^2}Str[Q^2_{i}]{\bf
J}_{2}(\mu)+{k_{i}\over 48\pi^3\mu^2}Str[{\bf 1}]{\bf J}_{3}(\mu)
\label{446}
\ee
\be
I_{3}^{massless}=-{k_{i}\over 64\pi^3\mu^2}Str[{\bf 1}]{\bf
J}_{3}(\mu)
\label{447}
\ee
Here
\be
{\bf J}_{n}\equiv {1\over V(\mu)}\int_{\cal F}{d^2\t\over
\im^n}X'(\mu)
\label{431}
\ee
which can be evaluated to be
\be
{\bf J}_{1}(\mu)=2\pi\log\mu^2 +2\pi(\log\pi+\gamma_{E}-3+{3\over
2}\log 3) +{\cal O}(e^{-{1\over \mu^2}})
\label{442}
\ee
\be
{\bf J}_{2}(\mu)=-{4\pi^2\over 3}(1+\mu^2)\;\;,\;\;
{\bf J}_{3}(\mu)=-\pi\log 3-{28\pi^3\over 15}\mu^4+{\cal
O}(e^{-{1\over \mu^2}})
\label{4431}
\ee
\def\m{\mu_{e}}

We would like now to describe the same calculation in the effective
field theory.

This calculation proceeds along the same lines as above taking into
account the following differences.

$\bullet$ Now $\m^2=1/k$ and $V(\m)=1/(8\pi\m^3)$.

$\bullet$ $\Gamma_{0}/V(\m)$ is given by the momentum mode part of
the stringy expression:
\be
{\Gamma_{0}\over V(\m)}=-4\m^3\partial_{\m^2}\sqrt{\im}\sum_{n\in Z}
e^{-\pi\im\m^2(2n+1)^2}
\label{450}
\ee

$\bullet$ There is an incomplete cancelation of the $1/8\pi\mu^2\im$
piece in
(\ref{F2}). What remains is $1/4\pi\im$.

$\bullet$ The integral over $\im$ is done from $0$ to $\infty$. We
will regulate the ultraviolet divergences by the Schwinger
regularization which amounts to integrating the parameter t in the
interval $[1/\pi\Lambda^2,\infty]$.

Then,
\be
{16\pi^2\over g_{i}^2}|_{1-loop}^{EFT}=L_{1}+L_{2}+L_{3}
\label{452}
\ee
where
\be
L_{1}=-{1\over \pi}Str\left[Q_{i}^2\left({1\over
12}-\chi^2\right)\right]{\bf K}_{1}(\m)
+{k_{i}\over 8\pi^2}Str\left[{1\over 12}-\chi^2\right]{\bf K}_{2}(\m)
\label{453}
\ee
\be
L_{2}=-{1\over 4\pi^2}\left(1+{1\over 3\m^2}\right)Str[Q^2_{i}]{\bf
K}_{2}(\m)+{k_{i}\over 16\pi^3}\left({1\over 2}+{1\over
3\m^2}\right)Str[{\bf 1}]{\bf K}_{3}(\m)
\label{455}
\ee
\be
L_{3}=-{k_{i}(1+2\m^2)\over 64\pi^3\m^2}Str[{\bf 1}]{\bf K}_{3}(\m)
\label{456}
\ee
and
\be
{\bf K}_{n}(\m)\equiv {1\over V(\m)}\int_{1\over
\pi\Lambda^2}^{\infty}{dt\over
t^n}\partial_{\m^2}\sqrt{t}\left[\th_{3}(it\m^2)-\th_{3}(4it\m^2)
\right]
\label{458}
\ee
The integrals can again be evaluated
\def\L{\Lambda}
\be
{\bf K}_{1}(\m,\L)=4\pi\log(\m/\L)+2\pi(\gamma_{E}-2)+
{\cal O}\left(e^{-\L^2/\m^2}\right)
\label{459}
\ee
and for $n>1$
\be
{\bf K}_{n}(\m,\L)=-{2\pi\over
n-1}\L^{2n-2}+8\pi^{2-n}(2n-3)(1-2^{2n-3})\Gamma(n-1)\zeta(2n-2)\m^{2n-2
}+
{\cal O}\left(e^{-\L^2/\m^2}\right)
\label{460}
\ee

In a similar fashion we can calculate the string one-loop correction
to the $R^2$ coupling with the result

\be
{1\over g_{R^2}^2}|_{1-loop} ={4\over \pi V(\mu)}\int_{\cal F}{d^2\t
\over \im|\eta|^4}
\sum_{a,b}^{1}\left[\partial_{\t}\left({\theta[^a_b]\over
\eta}\right)\left(\bar G_{2}-{1\over
6\mu^2}\partial_{\tb}\right)X'\right.
\label{48}\ee
$$\left.+{1\over 6\mu^2}{\theta[^a_b]\over \eta}\left(\bar
G_{2}-{1\over 6\mu^2}\partial_{\tb}\right)\partial_{\t}X'\right]+
{k(k+2)\over 16\pi V(\mu)}\int_{\cal F}{d^2\t\over \im^2}X'Z_{0}
$$
where
\be
\bar G_{2}\equiv \partial_{\tb}\log\bar\eta+{i\over 4\im}={1\over
2}\partial_{\tb}\log[\im\bar \eta^2]
\label{49}\ee

One-loop corrections to higher dimension operators can also be
computed.
We give here the result for $F_{\mu\nu}^4$ for $Z_{2}\times Z_{2}$
orbifold compactifications of the heterotic string. This correction
gets contributions from all sectors including $N=4$ ones and it is
thus interesting for studying decompactification problems in string
theory.
The $N=4$ sector contribution to the $F_{\mu\nu}^4$  term for the
$E_{8}$ gauge group can be computed from (\ref{F4}) to be
\be
{1\over g^2_{F^4}}|^{E_{8}}_{1-loop}={1\over V(\mu)}\int_{\cal
F}{d^2\t\over \im^2}
X'(\mu)\prod_{i=1}^{3}\left[\im\Gamma_{2,2}(T_{i},U_{i})\right]
\sum_{a,b=0}^{1}{\bar\vartheta^8[^a_b]\over \bar \eta^{24}}\times
\label{56}
\ee
$$
\times
\sum_{\gamma,\delta=0}^{1}\bar\vartheta^7[^{\gamma}_{\delta}]\left(
{i\over \pi}\partial_{\tb}-{5\over 2\pi\im}\right)\left({i\over
\pi}\partial_{\tb}-{1\over
4\pi\im}\right)\bar\vartheta[^{\gamma}_{\delta}]
$$

\section{IR Flow Equations for Couplings}
\setcounter{equation}{0}

Once we have obtained the one-loop corrections to the coupling
constants we can observe that they satisfy scaling type flows.
We will present here IR Flow Equations (IRFE) for differences of
gauge couplings.

The existence of IRFE is due to differential equations satisfied by
the
lattice sum of an arbitrary (d,d) lattice,
\be
Z_{d,d}={\rm Im}\tau^{d/2}\sum_{P_{L},P_{R}}e^{i\pi\tau
P^{2}_{L}/2-i\pi\bar\tau P_{R}^2/2}
\label{partition}
\ee
where
\be
P_{L,R}^2=\vec n G^{-1}\vec n+2\vec mBG^{-1}\vec n+\vec
m[G-BG^{-1}B]\vec m\pm 2\vec m \cdot \vec n
\label{momentum}
\ee
$\vec m,\vec n$ are integer d-dimesional vectors
and $G_{ij}$ ($ B_{ij}$) is a real symmetric (antisymmetric) matrix.
$Z_{d,d}$ is $O(d,d,Z)$ and modular invariant.
Moreover it satisfies the following second order differential
equation\footnote{The special case for $d=2$ of this equation was
noted and used in \cite{moduli,ant}.}:
\be
\left[ \left(G_{ij}{\partial\over \partial G_{ij}}+{1-d\over
2}\right)^2
+2G_{ik}
G_{jl}{\partial^2\over \partial B_{ij}\partial B_{kl}}
-{1\over 4}-4{\rm
Im}\tau^2{\partial^2\over \partial \tau\partial \bar
\tau}\right] Z_{d,d}=0\label{irfe}
\ee

The equation above involves also the modulus of the torus $\tau$.
Thus it can be used to convert the integrands for threshold
corrections to differences of coupling constants  into
total
derivatives on $\tau$-moduli space.
We will focus on gauge couplings of $Z_{2}\times Z_{2}$ orbifold
models.
To derive such an equation we start from the integral expressions of
such couplings (\ref{511})
to obtain
\be
\Delta_{AB}\equiv {16\pi^2\over g_{A}^{2}}-{16\pi^2\over g^{2}_{B}}=
-4\mu^3(b_{A}-b_{B})\int_{\cal F}{d^{2}\tau\over {\rm
Im}\tau^{2}}X'(\mu){\rm Im}\tau \Gamma_{2,2}(T,\bar T,U,\bar
U)\label{difere}
\ee
Eq. (\ref{difere}) does not apply to $U(1)$'s that can get enhanced
at special points of the moduli.
Using (\ref{irfe}) we obtain
\be
\left[\left(\mu{\partial \over \partial \mu}\right)^2-2\mu{\partial
\over \partial \mu}-16{\rm Im}T^2{\partial^2\over \partial T\partial
\bar T}\right]\Delta_{AB}=0\label{irfe2}
\ee
and we have also a similar one with  $T\to U$.
Note that for couplings that have a logarithmic behaviour, the double
derivative of $\mu$ does not contribute.

We strongly believe that such equations also exist for single
coupling constants using appropriate differential equations for
$(d,d+n)$ lattices.

Notice that the IR scale $\mu$ plays the role of the RG scale
in the effective
field theory:
\be
{16\pi^2\over g_{A}^{2}(\mu)}={16\pi^2\over
g_{A}^2(M_{str})}+b_{A}\log
{M^2_{str}\over \mu^2}+ F_{A}(T_{i})+{\cal O}(\mu^2/M^2_{str})
\label{betaa}
\ee
where the moduli $T_{i}$ have been rescaled
by $M_{str}$ so they are dimensionless.
Second, the IRFE gives a scaling relation for the moduli dependent
corrections.
Such relations are very useful for determining the moduli dependence
of the threshold corrections.
We will illustrate below such a determination, applicable to the
$Z_{2}\times Z_{2}$ example described  above.

Using the expansion (\ref{betaa}) and applying the IRFE (\ref{irfe2})
we obtain
\be
{\rm Im}T^2{\partial^2\over \partial T\partial\bar
T}(F_{A}-F_{B})={1\over 4}(b_{A}-b_{B})\label{eq2}
\ee
and a similar one for $U$.
This non-homogeneous equation has been obtained in \cite{moduli,ant}.

Solving them we obtain
\be
F_{A}-F_{B}=(b_{B}-b_{A})\log[{\rm Im}T{\rm Im}U]+f(T,U)+g(T,\bar
U)+{\rm cc}
\ee
If at special points in moduli space, the extra massless states are
uncharged
with respect to the gauge groups appearing in (\ref{eq2}) then the
functions
$f$ and $g$ are non-singular inside moduli space.
In such a case duality invariance of the threshold corrections
implies
that
\be
F_{A}-F_{B}=(b_{B}-b_{A})\log[{\rm Im}T{\rm
Im}U|\eta(T)\eta(U)|^4]+{\rm constant}
\ee
This is the result obtained via direct calculation in \cite{moduli}.

It is thus obvious that the IRFE provides a powerful tool in
evaluating
general threshold corrections as manifestly duality invariant
functions of the moduli.

\section{Further Directions}
\setcounter{equation}{0}

Another set of important couplings that we have not explicitly
addressed in this paper are the Yukawa couplings.
Physical Yukawa couplings depend on the k\"ahler potential and the
superpotential.
The superpotential receives no perturbative contributions and thus
can be calculated at tree level.
The Kh\"aler potential however does get renormalized so in order to
compute
the one-loop corrected Yukawa couplings we have to compute the
one-loop renormalization of the K\"ahler metric.
When the ground state has (spontaneously-broken) spacetime
supersymmetry the wavefunction renormalization of the scalars
$\phi_{i}$ are the same as those for their auxiliary fields $F_{i}$.
Thus we need to turn on non-trivial $F_{i}$, calculate their
effective action on the torus and pick the quadratic part
proportional to
$F_{i}\bar F_{\bar j}$.
This can be easily done using the techniques we developed in this
paper
since it turns out that the vertex operators \cite{atsen} for some
relevant $F$ fields are bilinears of left and right U(1) chiral
currents.

There are several open problems that need to be addressed in this
context.

The structure of higher loop corrections should be investigated.
A priori there is a potential problem, due to the dilaton, at higher
loops.
One would expect that since there is a region of spacetime where the
string coupling become arbitrarily strong, higher order computations
would be problematic. We think that this is not a problem in our
models, because
in Liouville models with N=4 superconformal symmetry (which is the
case we consider) there should be no divergence due to the dilaton at
higher loops.
However, this point need further study.
One should eventually analyze the validity of non-renormalization
theorems at higher loops \cite{ant} since they are of prime
importance for phenomenology.

The consequences of string threshold corrections for low energy
physics
should be studied in order to be able to make quantitative
predictions.

Finally, the response of string theory to the magnetic backgrounds
studied in this paper should be analysed since it may provide with
useful clues concerning
the behavior of strings in strong background fields and/or
singularities.

\vskip 1cm

\centerline{\bf Acknowledgements}

We would like to thank the organizers of the Spring School and
Workshop on String Theory for giving us the opportunity to present
our results.
E. Kiritsis would like to thank ICTP for its hospitality during the
workshop.
C. Kounnas  was  supported in part by EEC contracts
SC1$^*$-0394C and SC1$^*$-CT92-0789.

\end{document}